# A Generic Signaling Framework for Seamless Mobility in Heterogeneous Wireless Networks

Oliver Blume[1], Abigail Surtees[2], Ramon Aguero[3], Eranga Perera[4], and Kostas Pentikousis[5]

[1] Alcatel-Lucent Deutschland AG, Research & Innovation, D-70499 Stuttgart, Germany. E-mail: oliver.blume@alcatel-lucent.de
[2] Roke Manor Research Ltd, Romsey, Hampshire SO51 0ZN United Kingdom. E-mail: abigail.surtees@roke.co.uk
[3] University of Cantabria, Avda. los Castros s/n E-39005 Santander, Spain. E-mail: ramon@tlmat.unican.es
[4] National ICT Australia, Sydney, Australia. E-mail: Eranga.Perera@nicta.com.au
[5] VTT Technical Research Centre of Finland, Kaitoväylä 1, FI-90571 Oulu, Finland. E-mail: kostas.pentikousis@vtt.fi

*Abstract*—In recent years several wireless communication standards have been developed and more are expected, each with different scope in terms of spatial coverage, radio access capabilities, and mobility support. Heterogeneous networks combine multiple of these radio interfaces both in network infrastructure and in user equipment which requires a new multi-radio framework, enabling mobility and handover management for multiple RATs. The use of heterogeneous networks can capitalize on the overlapping coverage and allow user devices to take advantage of the fact that there are multiple radio interfaces. This paper presents the functional architecture for such a framework and proposes a generic signaling exchange applicable to a range of different handover management protocols that enables seamless mobility. The interworking of radio resource management, access selection and mobility management is defined in a generic and modular way, which is extensible for future protocols and standards.

I.  INTRODUCTION

Recent trends in telecommunications point to an increasing number of off-the-shelf devices with several integrated radio access technologies (RATs) and a variety of radio access offerings at the same location. For example, users can connect using a cellular network, wireless broadband MANs, and wireless city-wide LANs, or confederations thereof, based on inexpensive IEEE 802.11 access points. Yet, presently, such devices manage each access independently and, where they are combined, their combination is rather rudimentary: it is typically left to the user to decide which interface(s) to activate and then manually enable them. 3G/UMTS mobile phones are an exception, as they were designed to automatically fall back to GSM in the absence of 3G coverage, although this was mainly due to migration considerations, rather than network resource optimization. Besides the lack of "automation" in detecting and using multiple radio networks, another impediment is the lack of sophisticated support for simultaneous access.

The proliferation of wireless devices has also increased the demand for anywhere/anytime network service availability. Users are accustomed, for example, to maintaining cell phone conversations while moving, often at high speed; and so have similar expectations for data, video and other services. With multiple RATs available, one might expect that it would be easier to stay connected. However, while in a single-RAT, single-service paradigm (say, 3G/UMTS and voice communication) remaining connected (including handovers) is RAT–specific issue; in an IP-based, multi-service, multi-RAT paradigm, it is maintaining IP connectivity that is of essence, which is an inter-RAT issue.

Of course, each RAT can define its own mechanism for detecting and handing over to other RATs. The topic of the so-called vertical handovers is of great interest with extensive published literature. However, defining new mechanisms for each emerging RAT can lead to extremely complex interactions between different sets of protocols associated with each RAT type. Moreover, each pair of RATs would need to be considered separately when performing vertical handovers from one RAT to another. In short, seamless multiaccess with full mobility support and adequate network management facilities still evades the telecommunications industry.

The need for overall multi radio resource management [1], handover management (including supporting multiple mechanisms [2]) and path selection [3] functionality have been identified and discussed. This paper proposes a generic signaling framework which facilitates seamless roaming in heterogeneous radio access networks and can be customized for different environments. We discuss the details that govern the interaction between these mobility control functional entities, providing a consistent abstraction between different RATs. The rest of the paper is structured as follows. Section II reviews related work and sets the background for the proposed framework. Section III presents the architecture for mobility control and Section IV the crux of the paper: the generic signaling framework. Section V discusses our results and future outlook; Section VI concludes the paper.

II.  BACKGROUND AND RELATED WORK

The work presented in this paper is undertaken within the context of the Ambient Networks project (AN), which introduced the concept of an overlay network control layer called Ambient Control Space (ACS), detailed in [4] and [5]. The holistic approach taken by AN allows the consideration of not only radio-centric aspects, such as metrics accessible at the radio level, but also other resource requirements from the entire protocol stack and throughout the AN. Although this paper focuses on the interaction between the functional entities (FEs) involved in communication establishment and handover procedures, the ACS spans over a broader set of functionalities, including, for example, context awareness, session management, security and privacy (Fig. 1, which is based on [6]). The generic framework presented in this paper benefits from the background functionalities provided by the ACS. A key example is the on-demand establishment of dynamic agreements between networks using the composition framework [7], easing the process of information sharing between the entities involved. Such information sharing can be of great value when making decisions about which is the optimal access.

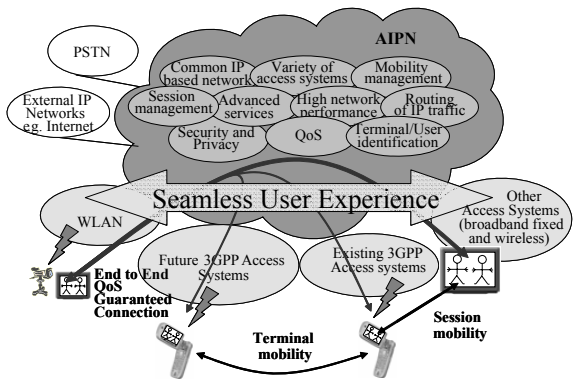

Fig. 1.  Control plane functions involved in seamless mobility support in a heterogeneous all-IP network.

The heterogeneity of radio accesses is currently being addressed by standardization bodies, such as 3GPP [8] and IEEE 802.21[9]. The scope of the latter is media independent handovers (MIH) based on link layer and other network- related information, allowing upper layers to optimize handovers between heterogeneous media. Although the standard defines primitives in terms of information elements for handover initiation and preparation, it does not take into consideration aspects such as access selection and IP mobility protocol execution. Our generic signaling framework has a wider scope and takes into account all three phases of a handover: initiation, preparation, and execution. In addition, it can interact with all relevant ACS entities and can make use of IEEE 802.21 functionalities when they are available.

Besides AN, other research groups are also defining alternative architectures which deal with the requirements posed by the forthcoming wireless communications scenarios. The Common Radio Resource Management (CRRM) framework, originally proposed by the IST Everest project [1] is currently continuing within the IST AROMA project [10]. However, this work is being carried out under tight guidelines based on 3GPP reference scenarios limiting CRRM to only currently available RATs, such as UMTS and WLAN. The generic framework designed and developed by AN can - encompass any RAT type that may emerge in the future, integrating it in a coherent manner.

### III. MOBILITY CONTROL ARCHITECTURE

Mobility control in heterogeneous wireless networks comprises three distinct stages: (a) monitoring and detecting available radio accesses, (b) handover decision-making, and (c) handover execution. As mentioned earlier, in heterogeneous wireless networks the available radio accesses typically have different coverage areas, so user mobility may lead to "vertical" handovers from one RAT to another, which cannot be handled by link layer protocols alone. This section focuses on the distribution of work during these stages between three FEs, namely, *Multi-Radio Resource Management* (MRRM), *Handover and Locator Management* (HOLM), and *Path Selection* The functional architecture of these high-level FEs (each of these may be composed of smaller functions or distributed among several locations), is illustrated in Fig. 2. The Flow Management FE is not involved in handover, but is shown in Fig. 2 as it is involved in setting up communication in the first place.

#### A. Monitoring and Detection of Available Radio Accesses

Legacy radio resource management (RRM) monitors the performance of the active radio link. Monitoring other RATs is out of scope of RRM and requires the introduction of a MRRM entity. MRRM interacts with the generic link layer and possibly legacy RRM entities [11] and is responsible for producing the detected access set (DAS). The DAS is the result of a first-order filtering of all available accesses: for example, user and operator policies, lack of a suitable subscription, or lack of resources in the candidate access networks may rule out the use of certain detected accesses. MRRM essentially takes care of this first stage of mobility control. Once the DAS is available, the second stage, handover decision-making may commence.

#### B. Handover Decision-making

Several events can lead to a handover decision for a mobile device, such as new accesses being added to the DAS, changes of radio signal strength or changes in the QoS requirements from running applications. Upon such events, MRRM may need to perform access selection. Traditionally, this selection was based on the characteristics of the first hop, that is, the radio link to the access network. In current and future networks the bottleneck from an end-to-end perspective may actually be beyond the first hop, e.g. on a satellite, WiMAX, HSPA, or even the wired backhaul link. Since MRRM concentrates on the first radio hop metrics, it may not be aware of performance bottlenecks or other hurdles to end-to-end communication. Moreover, there may be constraints on selecting the most appropriate access in DAS that are out of scope of radio resource management. Examples include availability and characteristics of candidate end-to-end paths, operator and user policies, and security considerations.

As the concept of a path-sensitive access selection was presented elsewhere [3], we will only briefly describe the basics here due to space considerations. The Path Selection FE comprehensively handles end-to-end path information based on, for example, entries in the policies database, relevant context information, security functions, mobility tools, network and service level management, and flow management. This way, different FEs are responsible for managing connectivity to the radio access network (MRRM) and to the end-to-end peer (Path Selection). This functional split is chosen because many handover operations (link layer and localized IP mobility such as using proxy MIP) can occur without impacting the end-to-end path, however such handover operations are not always possible.

MRRM and Path Selection can be designed as specialized functionalities; the split minimizes the extent to which specific or sensitive information has to be shared between different elements and hides the details of the protocols used for path assessment within Path Selection. As radio ratings can change on a sub-second time scale it is advantageous to allow MRRM to decide on the final selection and on the timing of a handover from the previous active set to the new active set. During the handover decision-making stage, Path Selection calculates a rating metric [2] and delivers a rating for each access in DAS. Based on this and on the radio-based ranking, MRRM selects the candidate access set (CAS) and the active access set (AAS). MRRM calls upon HOLM for the execution of the handover.

*C. Handover Execution*

During the third stage, handover execution, the radio device configuration and radio connectivity have to change, locators are updated, network connectivity needs to be restored, and finally context issues (AAA, QoS) have to be taken care of. In the current state of the art, protocol layering limits the interaction of radio layer protocols with IP mobility protocols, causing significant interruption during movement detection, address configuration and location update procedures. In order to obtain the best user experience, all these steps need to be coordinated and synchronized. In AN, this is the responsibility of HOLM.

Upon receipt of a handover request from MRRM, HOLM selects a handover protocol, for example, Mobile IP or HIP, and other supporting protocols (such as context transfer tools) from the mobility toolbox [12],[13]. Although MRRM has chosen the access to use, there may still be multiple end-to-end paths available over this RAT and there may be multiple locators available for use as the source locator (e.g. MIP home address or the locator local to the chosen access network). HOLM interacts with Path Selection to make the final choice of path and locator. Locator discovery and allocation is concerned with acquiring a set of locators (e.g. a MIP care-of-address or a HIP locator) that can be used for a particular interface to communicate in the connected IP access network. This part of locator management may be connectivity-specific (e.g. DHCP in an IPv4 network, stateless address auto-configuration in IPv6). Traditionally, these steps are performed by the network layer without interaction with the radio layer (and other ACS FEs). Generic and efficient handover and mobility management requires control of the activity and timing of the locator management. Path Selection will use interface(s) to protocols and the operating system to control the locators and return them to HOLM.

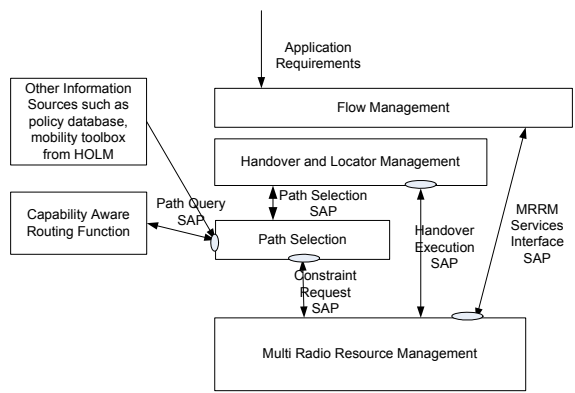

Fig. 2. Functional architecture and FEs in the mobility control plane

HOLM calls the handover protocol daemon to execute the change of locator. Further, HOLM commands MRRM to release and setup radio connectivity at the appropriate stage in the process. This is discussed further in Section IV. Mobility events should be invisible to flow management, but a (vertical) handover may change the characteristics of the end-to-end flow. As flow management requests radio resources from MRRM when an application flow is setup, it is up to MRRM to inform the Flow Management FE of such changes.

IV. MOBILITY CONTROL MESSAGE SEQUENCES

Fig. 2 shows the interaction between the FEs involved in our proposed mobility control architecture. The arrows between the FEs show *logical* communication, which may be node internal, occur across a radio link, or network interfaces. Communication may occur periodically or on an event-based manner and may involve one-to-one or one-to-many relationships. Therefore, the physical communication shall be supported by a communication tool, that enables event subscription and filtering and FE detection for bootstrapping, such as TRG [11],[14] or the IEEE 802.21 MIH Event Service [9].

The signaling sequence in Fig. 3 starts with MRRM scanning the environment, either periodically or in an event based manner. MRRM uses the functionality provided by other entities, such as the Generic Link Layer (GLL), which abstracts the particularities of the subjacent technologies, supplies a generic way of link setup and release, and accomplishes the scanning of potential accesses. As described above, MRRM generates the DAS and requests constraints from Path Selection. The response enables MRRM to narrow the set and finally define the AAS based on path and context constraints.

A vertical handover decision often requires the activation of a new radio interface in the mobile device. If the device can support simultaneous radio transmissions, this activation can be done without breaking the current connectivity (make-before-break handover, MBB HO). This capability will be known to MRRM. If MBB is not supported, then the old radio connection has to be broken and some interruption will occur (break-before-make handover, BBM HO). Therefore, preparative (or "backward") mobility protocols such as Fast Mobile IP (FMIP) [15] have been designed that perform certain time-consuming parts of the handover execution using the existing connection. This reduces the interruption time for BBM HO for devices with a single radio transmitter (or multiaccess devices than for whatever reason cannot employ more than one interface simultaneously and in a coordinated manner).

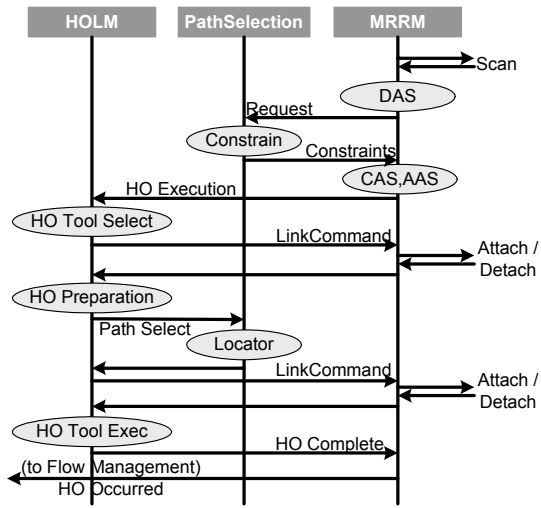

Fig. 3. Generic handover signaling sequence.
Link Commands can be sent by HOLM at any time.

From the above it is clear that the time for switching of the radio connectivity depends on the choice and performance of the handover tool. While for MBB and BBM the difference could be managed by MRRM, this is not the case, for example, for FMIP, because MRRM is not aware of the IP protocol steps. This implies cross layer communication between the radio link layer and the IP layer, that is, between MRRM and HOLM. A generic signalling sequence that applies for all HO tools is proposed in Fig. 3: the handover sequence is selected by HOLM and is transparent to the other FEs. HOLM controls the synchronisation between MRRM, HOLM and Path Selection by means of requests and responses. Link control commands can be sent to MRRM at any time of the sequence.

HOLM first selects the tool needed according to the characteristics posed by MRRM. Some handover tools will perform a preparation phase then. Next, Path Selection is called to choose the routing path. This may involve other functions (e.g. using the INQA [16] protocol or probing of the path). During this step, Path Selection chooses the locator from those provided in the target access. It returns the locator to HOLM which calls one of the handover protocol daemons to switch IP connectivity from the old to the new locator. HOLM confirms the handover execution to MRRM. In case the QoS performance changes due to the handover, MRRM informs the flow management FE, so that the services can be adapted.

Corresponding to the arrows in the signaling sequences is the exchange of some well defined primitives on the interfaces between HOLM, MRRM and Path Selection, as illustrated in Fig. 2. The main Service Access Points (SAPs) are:

(a) the *Constraint Selection SAP*, which is located between MRRM and Path Selection. It provides the primitives to request constraints on the access selection

(b) the *HO Execution SAP* located between MRRM and HOLM. It is used for the handover execution request and for link command requests during handover execution.

(c) the *Path Selection SAP* located between HOLM and Path Selection, and provides primitives for selection of path and retrieval of the locator for the IP protocol.

(d) the *MRRM Services Interface SAP* located between MRRM and Flow Management. It provides primitives for request and release of data flows with specified QoS parameters. Further, an indication for mobility related

TABLE I. LIST OF PRIMITIVES AND THEIR PARAMETERS ON THE INTERFACES BETWEEN HOLM, PATH SELECTION AND MRRM.

| SAP | FEs | Primitives | Parameters |
|---|---|---|---|
| Constraint Selection | MRRM PathSelect | Constraint Request Constraint Response | Flow ID, Cell IDs, Network IDs Cell IDs, Network IDs with rating |
| HO Execution | MRRM HOLM | HO Execution Request HO Complete LinkAttach & LinkSwitch Request LinkAttach & LinkSwitch Response LinkDetach Request LinkDetach Response | Flow ID, Current and Target Access, MBB Flag Result Flow ID, Current & Target Access, Requested QoS Result, Granted QoS Flow ID, Current Access Result |
| Path Query | HOLM PathSelect | PathSelect PathSelected | Flow ID, Target Access, FMIP Flag Result, New Locator |
| Flow Management | MRRM FlowMng | AccessFlowSetup AccessFlowSetup Response HandoverOccurred HandoverOccurred Response | Flow ID, Requested QoS Result, Granted QoS Flow ID, Provided QoS Result |

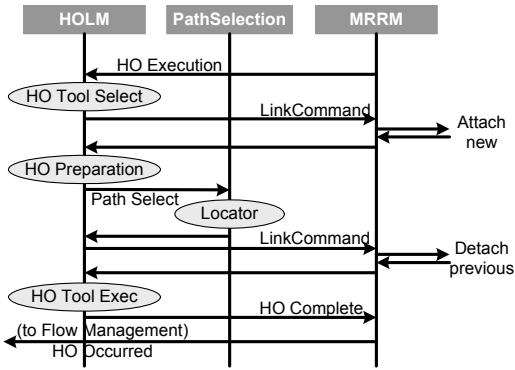

Fig. 4. Signaling sequence for a MBB HO

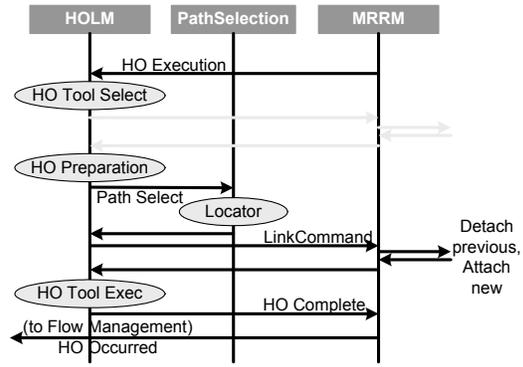

Fig. 6. Signaling sequence for a Fast Mobile IP HO

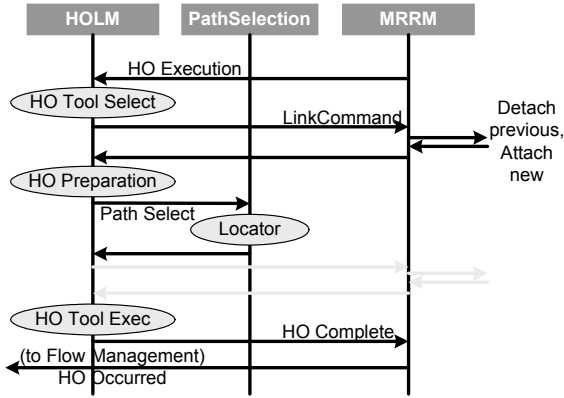

Fig. 5. Signaling sequence for a BBM HO

changes of QoS is provided. The primitives are listed in Table I.

In the following subsections we describe three different message exchange sequences for the MBB, BBM, and FMIP handover cases derived from the generic signaling framework described above and illustrated in Fig. 3. Each of the three corresponding cases begins with the handover request sent to HOLM and the differentiating factor is the timing of link control primitives.

### A. Make-Before-Make HO Signaling

In the *MBB scenario* (Fig. 4), HOLM establishes an attachment with the new access right after tool selection, without detaching from the former access. IP connectivity is established using multihoming in the mobile device. During HO execution the binding of the new locator is performed, e.g. a MIP Binding Update at the Home Agent. This switches the IP connectivity before finally the initial access is freed.

### B. Break-Before-Make HO Signaling

On the *BBM* alternative (Fig. 5), the former access has to be freed before attaching to the new one. Path Selection is called as soon as the new radio link is up, enabling the selection of the new locator. Finally, as before, the IP protocol daemon performs the update of locator binding.

### C. Fast Mobile IP HO Signaling

As can be seen, in the *Fast Mobile IP case* (Fig. 6), the sequence of actions is rather similar to the BBM case, although in this case the protocol daemon performs a preparation phase. During this a Proxy Router Advertisement is received in the mobile device which responds by sending a Fast Binding Update to the current access router, leading to interaction between current and target access routers for preauthentication and resource reservation. The success of this preparation phase is indicated to the FMIP protocol daemon in the mobile device with a Fast Binding Acknowledgement, which ends the HO preparation step of HOLM. Then, Path Selection can be called to select the new locator even before having attached to the new access. Now, the radio interface is switched from the old to the new access. FMIP provides a tunnel for forwarding data traffic to the previous access router, until finally HO execution is finalized by updating the locator (MIP Binding Update) by the protocol daemon. This minimizes the time in which the end user is not connected to any access.

## V. RESULTS AND OUTLOOK

In future heterogeneous wireless networks access selection, handover tool selection and handover execution cannot be handled independently or monolithically but require interaction of resource management, mobility management and QoS management functionality. This paper has proposed an architecture and generic signaling framework for seamless mobility in such networks. The interaction has been detailed for three different mobility protocols with message sequences and parameters.

Due to the general approach of the functional architecture the HO signaling sequence can also be applied to communication establishment, i.e. when a new communication session is started by the mobile user. This also involves access selection, locator selection and mobility protocol steps for locator binding, and the same primitives and sequences as for HO can be used, except that no previous location binding exist. Details of this are currently under development.

The correctness and applicability of the proposed generic signaling framework is currently being studied with a demonstrator implementation for seamless mobility in heterogeneous wireless networks. A first stage of implementation and integration of the ACS has been demonstrated recently at the IST summit 2007 with show cases including several mobility events. The update to a more complete set of features and functionality will be publicly demonstrated at the end of the AN project.

Last but not least, it is worth mentioning that within the 3GPP standardization work group SA2 [8], the vision of an all-IP network (AIPN, [6]) is pursued aiming at accommodating a variety of different access systems, and creating a multiaccess environment for SAE users. Mobility mechanisms will not depend on specific access, transport technology, or IP version. The specification process is now in "Stage 2", but mechanisms for access selection and handover management are still left "for future study". AN plans to contribute the work presented in this paper to 3GPP for standardization.

## VI. CONCLUSIONS

In this paper we focused on how the Ambient Networks framework can be used to improve current procedures on access selection and mobility management, highlighting the interaction between the MRRM and the HOLM FEs, based on a generic set of interfaces and the corresponding signaling. They bring about the possibility to take into consideration a broader range of metrics when deciding which access to select. In addition, different mobility protocols can coexist, providing the ability to select the one which better fits the requirements at a particular time.

One of the main drawbacks that can be attributed to how current networks are evolving is that each new aspect is handled on an isolated way, without looking on whether it might have some additional implications in other parts of the whole system. The result is a continuously growing patchwork of different techniques and protocols. The Ambient Networks project as a whole and this work in particular, aims, to break this tendency by proposing a generic, flexible framework, which can host different entities as needed, with the great added value that they are able to use all the functionalities which were already available within the ACS.

## ACKNOWLEDGMENT

This work has been carried out in the framework of the Ambient Networks project (IST 027662), which is partially funded by the Commission of the European Union. The views expressed in this paper are solely those of the authors and do not necessarily represent the views of their employers, the Ambient Networks project, or the Commission of the European Union. The comments and ideas from people involved in the project's multiaccess and mobility research are gratefully acknowledged.